\documentstyle[prl,aps,multicol]{revtex}
\tighten
\input psfig
\begin{document}
\draft
\title{
Quantum Coulomb Blockade in Chaotic Systems}
\author{I.L. Aleiner$^{1}$ and L.I. Glazman$^{2}$}
\address{$^{1}$NEC Research Institute,
4 Independence Way, Princeton, NJ 08540\\ $^{2}$Theoretical Physics Institute,
University of Minnesota, Minneapolis MN 55455}
\maketitle

\begin{abstract}
We study the Coulomb blockade in a chaotic cavity connected to a lead
by a perfectly transmitting quantum channel. In contrast to the
previous theories, we show that the quantum fluctuations of charge,
resulting from the perfect transmission, do not destroy the Coulomb
blockade completely.  The oscillatory dependence of all the observable
characteristics on the gate voltage persists, its period is not
affected; however, phases of the oscillations are random, reflecting
the chaotic electron motion in the cavity. Because of this randomness,
the Coulomb blockade shows up not in the averages but in the
correlation functions of the fluctuating observables ({\em e.g.},
capacitance or tunneling conductance).
\end{abstract}

\begin{multicols}{2}
The effect of Coulomb blockade in chaotic systems sets up an ideal
stage for studying the interplay between the quantum chaos and
interaction phenomena in a many-electron system.  By tuning the
connection between the leads and chaotic cavity, Fig.~\ref{cavity},
one can study a rich variety of non-trivial effects. In the weak
tunneling limit, discrete charging of the cavity results in a system
of sharp conductance peaks, which carry information about the chaotic
motion of non-interacting electrons confined inside an almost closed
cavity. In the opposite limit of wide channels, charge quantization
does not occur, and quantum chaos of free electrons in an open
billiard may be studied. In a broad intermediate region, the charge
quantization is gradually destroyed, and the chaotic electron motion
is affected by fluctuations of charge of the cavity. The modern
experimental technique\cite{Chang,Marcus} allows one to continuously
traverse between these regimes.

The effect of charging is described by means of the  Hamiltonian\cite{Review} 
\begin{equation}
\hat{H}_C=\frac{E_C}{2}\left(\frac{\hat{Q}}{e}-{\cal N}\right)^2,
\label{Hc}
\end{equation}
where the conventional parameter ${\cal N}$ is related to the gate
voltage $V_g$ and gate capacitance $C_g$ by ${\cal N}=V_g/eC_g$, and
$\hat{Q}$ is the cavity charge. Charging energy $E_C$ for large
cavities is much larger than the mean level spacing $\Delta$. If the
connection of the cavity with the leads is weak, the charge is well
quantized for almost all $\cal N$ except narrow vicinities of the
charge degeneracy points (half-integer ${\cal N}$). The behavior of
the differential capacitance of the cavity, $d\langle Q\rangle/dV_g$
and of the conductance through the cavity is quite different for the
system tuned to the immediate vicinity of charge degeneracy points
(Coulomb blockade peaks), or away from these points (Coulomb blockade
valleys). The statistics of the peaks can be related to the properties
of  single electron energies and wave functions\cite{Stone}, so that the
distribution functions for observable quantities can be extracted from the
well known random matrices results. The transport in the valleys
occurs by virtual transitions of an electron via excited states of the
cavity\cite{Averin90}. The statistics of  conductance in this case
was recently obtained in Ref.~\cite{Aleiner96}.

{\narrowtext
\begin{figure}[h]
\vspace{-1.42cm}
\hspace*{-0.2cm}
\psfig{figure=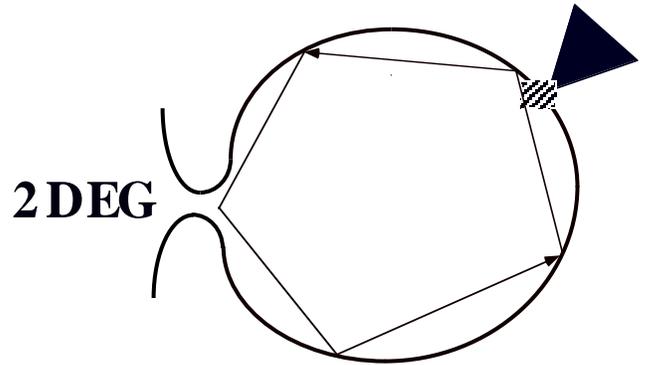,height=13cm}
\vspace{-6.63cm}
\caption{Schematic view of a chaotic cavity connected to a lead by an ideal
quantum channel. Second lead (triangle) can be attached to the cavity by weak
tunnel junction for measurements of two-terminal conductance.} 
\vspace{-0.22cm}
\label{cavity}
\end{figure}
}
 
All the aforementioned results are obtained neglecting quatum
fluctuations of the charge of the cavity. These fluctuations become
large with increase of the coupling between the cavity and the
lead. Then, the difference between the peaks and valleys becomes less
pronounced and eventually instead of the peak structure, one observes
only a weak periodic modulation. This situation can be described
neither by the properites of the single-electron wave function nor by
the lowest order virtual transitions via the excited states.  The case
of almost perfect transmission of a one-channel point contact
connecting a cavity with the lead was analysed recently by
Matveev\cite{Matveev95}. According to Ref.~\cite{Matveev95}, the
Coulomb blockade disappears completely if the transmission coefficient
of the point contact is exactly unity and $\Delta = 0$. The strength
of the theory\cite{Matveev95} is in non-perturbative treatment of the
Coulomb interaction; the drawback is in neglecting the chaotic motion
of electrons in the cavity.

In this Letter, we account for both the strong quantum charge
fluctuations, and the chaotic electron motion within the cavity. As we
will see, backscattering of electrons from the walls of the cavity
into the channel connecting it to the lead results in oscillations of
observables with the gate voltage. In the limit of perfect channel
transmission, the relative magnitude of the differential capacitance
oscillations is $\sqrt{\Delta/E_C}$ and
$\left(\Delta/E_C\right)\ln^2\left(E_C/\Delta\right)$ for the spinless
and for spin $1/2$ cases respectively. If the second lead is attached
to the cavity (see Fig.~\ref{cavity}) by a weak tunnel junction with
conductance $G_0 \ll e^2/\hbar$, the two-terminal conductance $G$
can be measured. The average value of the conductance is supressed by
the Coulomb blockade, $\langle G\rangle \simeq G_0\Delta/E_C$,
fluctuations of the conductance are of the order of its average. This
resembles the result for the elastic cotunneling in weak coupling
regime\cite{Averin90,Aleiner96}.  However, the distinction between
valleys and sharp peaks is lost.

{\em Spinless electrons, qualitative discussion} -- Here, we use an
analogy to the Fermi liquid theory of the Kondo problem due to
Nozieres\cite{Nozieres}.  The dynamics of the system at energies
smaller than $E_C$ can be well described by the lead and the cavity
effectively separated from each other.  Indeed, an electron inside the
cavity is not affected by the interaction, unless, in the course of
chaotic motion, it encounters the channel entrance. Even if the
trajectory reaches the channel entrance (Fig.~\ref{cavity}), the
electron is scattered back, provided its energy $\epsilon$ (measured
from the Fermi level) is much smaller than the charging energy
$E_C$. The phase shift $\delta$ of this scattering at $\epsilon =0$ is
fixed by the Friedel sum rule $\delta(0) =\pi{\cal N}$, where we used
the fact that for an ideal transmitting channel the charge of the
cavity is $\langle Q\rangle=e{\cal N}$\cite{Matveev95}. On the other
hand, the variation of the phase of the electron reflection from the
channel entrance $\alpha(\epsilon/E_c)= \delta (\epsilon,{\cal
N})-\pi{\cal N}$ is of the order of unity in the energy interval
$|\epsilon|<E_C$.

Because an electron at low energy $\epsilon$ can not leave the cavity, its
motion in the cavity is quantized.  The one-electron energies are  periodic
functions of ${\cal N}$, with characteristic amplitude $\Delta$ and random
phases (the latter assumption will be further elaborated on).  The oscillating
with
${\cal N}$ contribution to the ground state energy $\delta { E}({\cal N})$ is
determined by sum of the eigenvalues in the energy interval $|\epsilon|<E_C$;
therefore $\delta { E}({\cal N})$ is the sum of  approximately
$E_C/\Delta \gg 1$ random numbers. As the result, the average
$\langle\langle\delta E({\cal N})\rangle\rangle$ vanishes, and the correlation
function of the differential capacitance ( in units of
$C_g$)
\begin{equation}
K_C({\cal N}-{\cal M})\equiv\frac{1}{E_C^2}
\partial^2_{{\cal M}}\partial^2_{{\cal N}}\langle\langle\delta E({\cal
N})\delta E({\cal M})\rangle\rangle
\label{difcap}
\end{equation}
takes the form
\begin{equation}
K_C ({\cal N}_-) =  \Lambda(0) \left(\frac{\Delta}{E_C}\right)
\cos 2\pi{\cal N}_-,
\label{r1}
\end{equation}
where the coefficient $\Lambda(0)$, see Eq.~(\ref{Bfunction}), is the
result of the effective action theory outlined later.

In order to explain the functional form of $K_C$ in Eq.~(\ref{r1}) and
make our argumentation more precise, it is very instructive to
evaluate the shift of the energy levels starting from the Gutzwiller
trace formula\cite{Gutzwiller}.  The energy of the ground state is
given by $E=-\int_{-\infty}^{0} d \epsilon N(\epsilon)$, where
$\epsilon$ is measured from the Fermi level, and the integrated
density of states $N(\epsilon)=\sum_i\theta(\epsilon -\epsilon_i)$ is
expressed as a sum over the classical periodic orbits:
\begin{equation}
N(\epsilon) ={\rm Re} \sum_j R_j(\epsilon)
\exp\left[\frac{i}{\hbar}{S_j(\epsilon)} +2in_j\delta(\epsilon,{\cal
N})\right].
\label{Gutzwiller}
\end{equation}
Here $R_j$ is the weight associated with $j$-th orbit, $S_j$ is the reduced
action for this orbit. The last term in the exponent in Eq.~(\ref{Gutzwiller})
characterizes the reflection from the entrance of the cavity, and 
$n_j$ is the number of such reflections for $j$th orbit. We omitted the mean
value of $N(\epsilon)$ which is independent of the phase shift $\delta$. 
In the double sum over the periodic
orbits, arising in evaluation of (\ref{difcap}),  one can retain only diagonal
terms\cite{Berry85} because  different orbits have 
different actions; the non-diagonal terms oscillate strongly and
vanish upon the averaging. We obtain from Eqs.~(\ref{Gutzwiller}) and 
(\ref{difcap})
\begin{eqnarray}
K_C({\cal N}_-)&\simeq& {\rm Re}\!\!\int_{-\infty}^0\!\!
\frac{\!d\epsilon_1 d\epsilon_2}{E_C^2} \sum_j\!\left|R_j\right|^2
n_j^4 e^{2\pi i n_j{\cal
N}_-}e^{\frac{i}{\hbar}(\epsilon_1-\epsilon_2)t_j}\nonumber\\ &&\times
e^{2in_j[\alpha(\epsilon_1/E_C) -\alpha(\epsilon_2/E_C)]},
\label{diagonal}
\end{eqnarray}
where we used the expansion $S_j(\epsilon_1)
-S_j(\epsilon_2) = (\epsilon_1-\epsilon_2)t_j$, with $t_j$ being the period of
$j$-th orbit. 

In a chaotic system the orbits proliferate exponentially with the
increasing period and cover all the energy shell.  This exponential
proliferation is compensated by the damping of the weight
$\left|R_j\right|^2$ which eventually leads to the classical sum
rule\cite{sumrule}
\begin{equation}
\sum_j\left|R_j\right|^2\dots \to \frac{1}{2\pi^2}\int_0^\infty
\frac{dt}{t}\dots.
\label{sumrule}
\end{equation}
valid for periods $t$ much larger than the period of the
shortest orbit $\hbar/E_T$. Energy $E_T$ associated with the time
scale at which the classical dynamics becomes ergodic is the
counterpart of the Thouless energy for the diffusive system.
Typically $E_T \gtrsim E_C$, therefore we adopt approximation $E_T
\gg E_C\gg\Delta$.  Application of Eq.~(\ref{sumrule}) to
Eq.~(\ref{diagonal}) yields
\begin{equation}
K_C({\cal N}_-)\simeq\!\!\int_0^\infty\! \frac{dt}{t}
F\left(\frac{E_Ct}{\hbar}\right)\sum_n n^4 P_n(t) 
\cos {2\pi n{\cal N}_-}.
\label{diagonal1}
\end{equation}
Here $F\left(\frac{E_Ct}{\hbar}\right)=\left|\int_{-\infty}^0
\frac{\!d\epsilon}{E_C} e^{\frac{i}{\hbar}\epsilon t+
2i\alpha(\epsilon/E_C) }\right|^2 $ is a dimensionless function. The
number of long orbits is exponentially large and they uniformly cover
the available phase space, therefore, we can employ the stastical
description of the reflections by the entrance of the cavity. The
probability for an orbit of period $t\gg \hbar/\Delta$ to reach the
entrance even once is small, $P_1(T)=t\Delta/\hbar$, and one can
neglect the orbits encountering such a reflection more than once. With
this simplification, we find Eq.~(\ref{r1}) with
$\Lambda(0)=\int_0^\infty dx F(x)$.  The dimensionless integral here
converges at $x\sim 1$, which corresponds to the orbits with $t\sim
\hbar/E_C\ll\hbar/\Delta$ and justifies the neglection with the terms
$n\neq 1$ in Eq.~(\ref{diagonal1}). The numerical coefficient given by
this integral depends on the phase shift $\alpha (\epsilon/E_C)$, and
cannot be found within the simple consideration.

Finally, we discuss the correlation of the differential capacitances
as a function of magnetic field. The magnetic flux threading a
periodic orbit adds a phase $\phi=A_jH/\Phi_0$ to the action in each
term $j$ in the Gutzwiller formula (\ref{Gutzwiller}), where $A_j$ is
the directed area under the trajectory, $H$ is the applied magnetic
field, and $\Phi_0$ is the flux quantum. Correspondingly, each term in
the sum (\ref{diagonal}) acquires factor $\cos (B_1A_j/\Phi_0)\cos
(B_2A_j/\Phi_0)$. In a chaotic system, $A_j$ is a random quantity
which fluctuations grow linearly with the period of the orbit,
$\langle\langle A_j^2(t)\rangle\rangle\simeq (E_Tt/\hbar){\cal A}^2$,
where ${\cal A}$ is a shape-dependent area proportional to the area of
the cavity.  Hence, the contributions from the trajectories with large
period, $\langle\langle A_j^2(t)\rangle\rangle \gtrsim
\Phi_0^2/(B_1\pm B_2)^2$ become exponentially suppressed. On the
other hand, the main contribution to the capacitance fluctuations
comes from the orbits with the period $t \lesssim\hbar/ E_C$. Thus,
the correlation magnetic field $B_c$ is controlled by the energy
$E_C$:
\begin{equation}
B_c=\frac{\Phi_0}{\cal A}\sqrt{\frac{E_C}{2\pi E_T}}.
\label{Bc}
\end{equation}
The exact formula for the correlation function
\begin{eqnarray}
&&K_C({\cal N}; B_1,B_2)=
\frac{\Delta}{2E_C}\left[
\Lambda\!\left(\frac{B_+^2}{B_c^2}\right)
+
\Lambda\!\left(\frac{B_-^2}{B_c^2}\right)
\right]\cos 2\pi{\cal N},  \nonumber\\
&&\Lambda (z) = \int_0^\infty\frac{dy}{y^2}
\exp \left(-2\int_0^\infty \frac{dx e^{-xy}}{x+1}\right)e^{-zy}
\label{Bfunction}
\end{eqnarray}
and numerical coefficient in Eq.~(\ref{Bc}) requires a rigorous theory
which we outline below. In Eq.~(\ref{Bfunction}), we introduced the
short hand notation $B_\pm=B_1\pm B_2$.

{\em Effective action theory} --
We start from the Hamiltonian of the system
\begin{equation}
\hat{H}=\hat{H}_{F} + \hat{H}_C,
\label{Hamiltonian}
\end{equation} 
where $\hat{H}_{F}$ is the Hamiltonian of non-interacting electrons in
the cavity and in the lead and the Coulomb blockade type interaction
$\hat{H}_C$ is described by Eq.~(\ref{Hc}).

We calculate the thermodynamic potential 
$
\Omega({\cal N}) = - T\ln {\cal Z},\
{\cal Z} = {\rm Tr} e^{-\beta {\hat H}},
$
where $T=1/\beta \ll E_C$ is the temperature of the system. The second
derivative  $\partial_{\cal N}^2 \Omega({\cal N})$ 
gives the differential capacitance of the system. 
Because the interaction $\hat{H}_C$ does not affect the electron dynamics 
inside the cavity, we can integrate out the fermionic degrees of freedom 
of the cavity and obtain a formally exact expression\cite{Aleiner97} for the
partition  function
\begin{equation}
{\cal Z} = e^{-\beta\Omega_0}\langle T_{\tau}e^{-\hat{S}} \rangle,
\label{Z}
\end{equation}
where the ${\cal N}$-independent energy $\Omega_0$ does not
contribute to the differential capacitance and can be neglected, and
$T_{\tau}$ stands for the imaginary time ordering. The averaging
$\langle\dots\rangle$ is performed over the effective one-dimensional
Hamiltonian \cite{Matveev95}
\begin{eqnarray}
\hat{H}_0({\cal N}) &=& 
iv_F\int dx \left\{\psi_L^\dagger\partial_x\psi_L -
\psi_R^\dagger\partial_x\psi_R
\right\} \label{Heff}\\
&&+ \frac{E_C}{2}\left(\int_{-\infty}^0dx
:\psi_L^\dagger\psi_L +\psi^\dagger_R\psi_R:+{\cal N}\right)^2,
\nonumber
\end{eqnarray}
where $v_F$ is the Fermi velocity in the channel, and $:\dots :$ stands for the
normal  ordering. The effective action $\hat{S}$ is
\begin{equation}
\hat{S}=\int_0^\beta d\tau_1 d\tau_2
L\left(\tau_1-\tau_2\right)\bar{\psi}(\tau_1) {\psi}(\tau_2),
\label{action}
\end{equation}
where $\psi(\tau)=e^{\tau\hat{H}_0}\left(\psi_L(x=0)+
\psi_R(x=0)\right)e^{-\tau\hat{H}_0}$ are the fermionic operators in 
the Matsubara representation, $\bar{\psi}(\tau) = \psi^\dagger(-\tau)$. 
The kernel $L(\tau)$ can be expressed in terms of the exact Green function of 
the closed cavity:
\begin{equation}
L(\tau)=\!
\frac{1}{4m^2}\!
\int\! dydy^\prime
\phi({y})\phi({y}^\prime)
\partial^2_{xx^\prime}
\tilde{{\cal G}}(\tau; {\bf r}, {\bf r}^\prime),
\label{kernel}
\end{equation}
where $\tilde{\cal G}={\cal G}-{\cal G}_\infty$, the exact Matsubara Green
function ${\cal G}$ is defined with zero boundary conditions (closed cavity), 
${\cal G}_\infty$ is the Green function corresponding to
an infinite ballistic cavity, and $\phi(y)$ is the wave
function of the transverse motion in the contact.  Physically, kernel
(\ref{kernel}) accounts for all possible returns of an electron to the
contact during its chaotic motion within the cavity.

If conductance of the junction connecting the cavity to the second
lead is small, $G_0 \ll e^2/\pi\hbar$, its effect can be accounted for
in the second order perturbation theory. Derivation similar to that of
Eqs.~(\ref{Heff})-(\ref{kernel}) yields for the two-terminal conductance
\begin{equation}
\frac{G}{G_0}= \frac{1}{\pi\nu V}{\rm Im}\left[\lim_{i\Omega_n\to
V+i0}\int_0^\beta d\tau \frac{\pi Te^{-i\Omega_n\tau}}{\sinh \pi
T\tau}\Pi(\tau)\right],
\label{Kuboformula}
\end{equation}
where $\nu$ is the averaged density of states per unit area, 
$\Omega_n=2\pi Tn$ is the bosonic Matsubara frequency, and function
$\Pi(\tau)$ is given by
\begin{eqnarray}
\Pi (\tau )&=&\frac{1}{\langle T_\tau e^{-\hat{S}}\rangle}\int_0^\beta d\tau_1
d\tau_2 L_e(\tau_1)L_e^*(\tau_2) \label{Pitau}\\
&&\times\langle T_\tau e^{-\hat{S}}
\hat{\bar{F}}(\tau)\hat{F}(0)\bar{\psi}(\tau -\tau_1)\psi (\tau_2)
\rangle.
\nonumber
\end{eqnarray}
In Eq.~(\ref{Pitau}), we retained only the contribution non-vanishing
in the limit $T \to 0$.  Unitary operator $\hat{F}$ in
Eq.~(\ref{Pitau}) shifts the number of electrons in the cavity by
one\cite{Furusaki95}, $\hat{F}^\dagger \hat{H}_0({\cal N}) \hat{F}=
\hat{H}_0({\cal N}+1)$; the kernel $L_e(\tau)$ describes the motion of
an electron from the tunneling junction ${\bf r}_t$ to the entrance of
the cavity, and is related to the exact one-electron Green function of
the closed cavity:
\begin{equation}
L_e(\tau)=
\frac{1}{2m}
\int dy
\phi({y})
\partial_{x}
{{\cal G}}(\tau; {\bf r}_t, {\bf r}).
\label{Le}
\end{equation}
In the case of electrons with spins, we imply summation over the spin
indices in Eqs.~(\ref{Heff}), (\ref{action}) and (\ref{Pitau}). 

The advantage of the representation (\ref{action}) and (\ref{Pitau})
is that the Hamiltonian (\ref{Heff}) is exactly solvable by the
bosonization technique\cite{Matveev95}, and the action $S$ can be
treated perturbatively provided that $\Delta \ll E_C,E_T$.  Green
function ${\cal G}$ is a random quantity, and its statistics can be
obtained using the well-known expansion in the diffuson and Cooperon
modes\cite{AronovSharvin}. In the regime $E_C \ll E_T$ the results
become universal, i.e. independent on the details of chaotic dynamics
in the system.  Below, we present the results of the calculation;
details will be reported elsewhere\cite{Aleiner97}.

{\em Spinless electrons, results} -- For spinless electrons (Zeeman
splitting exceeds the Fermi energy of electrons in the channel), it
suffices to consider only the first order perturbation theory in
$\hat{S}$ for the calculation of the capacitance which gives
Eq.~(\ref{r1}). Conductance is found from Eq.~(\ref{Kuboformula}) and
Eq.~(\ref{Pitau}) neglecting action $S$ at all. For the average
conductance, we find
\begin{equation}
\langle\langle G\rangle\rangle =G_0\frac{\Delta}{\gamma^2E_C}\Lambda (0),
\label{spinlessaverage}
\end{equation}
where $\gamma\approx 1.78\dots$ is the Euler constant and function
$\Lambda(x)$is defined in Eq.~(\ref{Bfunction}). 
The correlation function of the conductance fluctuations is given by
\begin{equation}
\frac{\langle\langle{\delta G_1\delta G_2}\rangle\rangle}
{\langle\langle G \rangle\rangle^2}
\!=\!\left[
{\Lambda^2\left(\frac{B_+^2}{B_c^2}\right)} +
{\Lambda^2\left(\frac{B_-^2}{B_c^2}\right)}
\right]\! 
\frac{\cos^2\pi{\cal N}_-}{\Lambda^2(0)},
\label{spinlesscorr} 
\end{equation}
where the correlation magnetic field is given by Eq.~(\ref{Bc}),  we
introduced the short hand notation, $G_i = G({\cal N}_i, B_i)$,  and $B_\pm =
B_1\pm B_2$, ${\cal N}_-={\cal N}_1- {\cal N}_2$.

{\em Spin $1/2$ case, results} --
In the spin $1/2$ case, the non-vanishing correlation function of the
differential capacitance appears only in  the second order of perturbation
theory in $\hat S$. In zero magnetic field, we find
\begin{equation}
K_C^{\sigma}({\cal
N}_-)=\frac{2\Delta^2}{3\pi^2E_C^2}\ln^4\left(\frac{E_C}{T}\right)
\cos 2\pi {\cal N}_-.
\label{energyspin}
\end{equation}
We present here also the correlation function in the unitary limit 
($B_+\gtrsim B_c$),
\begin{equation}
K_C^{\sigma}=
\frac{16\Delta^2}{3\pi^2E_C^2}\ln^3\left|\frac{B_c}{B_-}\right|\ln
\left|\frac{E_C}{T}\left(\frac{B_-}{B_c}\right)^{\frac{3}{2}}\right|
\cos 2\pi {\cal N}_-,
\label{energyspinu}
\end{equation}
valid for the region of fields $B_c\sqrt{T/E_C}\ll B_-\ll B_c$.

To calculate the average conductance, one can neglect $\hat{S}$, similarly to
the spinless case. This yields
\begin{equation}
\langle\langle G\rangle\rangle=G_0\frac{\Delta}{\gamma E_C}
\ln\left(\frac{E_C}{T}\right).
\label{spinaverage}
\end{equation}
The main conribution to the conductance correlation function is independent
on ${\cal N}_-$, but depends on the magnetic field; in the unitary limit
\begin{equation}
\frac{\langle\langle{\delta G_1\delta G_2}\rangle\rangle}
{\langle\langle G\rangle\rangle^2}=\frac{1}{2}
\left[\ln\left(\frac{B_-}{B_c}\right)/\ln\left(\frac{E_C}{T}\right)\right]^2.
\label{spincorrb}
\end{equation}
The characteristic magnitude of oscillating with ${\cal N}$ contribution to the
correlation function is smaller than Eq.~(\ref{spincorrb}) by a parameter
$\Delta/E_C$.

At low temperatures $T$, the above results diverge, which indicates that the
higher order terms in $\hat{S}$ should be taken into account. Using a
variational approach\cite{Aleiner97}, we  conclude that $T$ should be replaced
with $\Delta\ln^2(E_C/\Delta)$ in Eqs.~(\ref{energyspin})-(\ref{spincorrb}).

In conclusion, Coulomb blockade oscillations persist even if a cavity
is connected to a lead by a {\em perfect} single-mode channel. This
effect is ignored if one approximates the electron spectrum in the
cavity by a continuum\cite{Matveev95}, or neglects the charge
quantization\cite{Buettiker96}.

We are grateful to B.L. Altshuler, A.I. Larkin and K.A. Matveev for useful
discussions. Hospitality of the Aspen Center for Physics is also acknowledged.
This work was supported by NSF Grant DMR-9423244.

\end{multicols} 

\begin{references} 
\bibitem{Chang} A.M. Chang {\it et. al.}, Phys. Rev. Lett. {\bf 76}, 1695
(1996). 
\bibitem{Marcus} J.A. Folk {\it et. al.}, Phys. Rev. Lett. {\bf
76}, 1699 (1996).  
\bibitem{Review} See {\it e.g.} D.V. Averin and K.K.  Likharev in {\it
Mesoscopic phenomena in solids}, edited by B.L. Altshuler, P.A. Lee, and
R.A. Webb (North Holland, New York, 1991).
\bibitem{Stone} R. Jalabert, A.D. Stone, and Y.
Alhassid, Phys. Rev.  Lett. {\bf 68}, 3468 (1992); Y. Alhassid and H. Attias,
{\em ibid}.  {\bf  76}, 1711 (1996). 
\bibitem{Averin90} D.V. Averin and Yu.N. Nazarov, Phys. Rev. Lett.
{\bf 65}, 2446 (1990).
\bibitem{Aleiner96} I.L. Aleiner and L.I. Glazman, Phys. Rev. Lett.
{\bf 77}, 2057 (1996).
\bibitem{Matveev95} K. A. Matveev, Phys. Rev. B {\bf 51}, 1743 (1995).
\bibitem{Nozieres}P. Nozieres, J. Low Temp. Phys., {\bf 17}, 31 (1974). 
\bibitem{Gutzwiller} M.C.  Gutzwiller, {\em Chaos in Classical and Quantum
Mechanics}, (Springer-Verlag, New York, 1990).
\bibitem{Berry85}M.V. Berry, Proc. R. Soc. London A, {\bf 400}, 229 (1985).
\bibitem{sumrule} J.H. Hannay and A.M.O. de Almeida, J. Phys. A {\bf 17}, 3429
(1984). 
\bibitem{Aleiner97} I.L. Aleiner and L.I. Glazman (in preparation).
\bibitem{AronovSharvin} See {\it e.g.} A.G. Aronov and Y.V. Sharvin, Rev.
Mod. Phys. {\bf 59}, 755 (1987) and references therein. 
\bibitem{Furusaki95} A. Furusaki and K.A. Matveev, Phys. Rev. Lett. {\bf 
75}, 709 (1995); Phys. Rev. B {\bf 52}, 16676 (1995).
\bibitem{Buettiker96} V.A. Gopar, P.A. Mello, M. B\"uttiker,
Phys. Rev. Lett. {\bf 77}, 3005 (1996).
\end{references}
\end{document}